# Image Enhancement with Statistical Estimation

Aroop Mukherjee[1] and Soumen Kanrar[2]

[1,2]Member of IEEE
*mukherjee_aroop@yahoo.com*
[2]Vehere Interactive Pvt Ltd Calcutta India
*Soumen.kanrar@veheretech.com*

## ABSTRACT

*Contrast enhancement is an important area of research for the image analysis. Over the decade, the researcher worked on this domain to develop an efficient and adequate algorithm. The proposed method will enhance the contrast of image using Binarization method with the help of Maximum Likelihood Estimation (MLE). The paper aims to enhance the image contrast of bimodal and multi-modal images. The proposed methodology use to collect mathematical information retrieves from the image. In this paper, we are using binarization method that generates the desired histogram by separating image nodes. It generates the enhanced image using histogram specification with binarization method. The proposed method has showed an improvement in the image contrast enhancement compare with the other image.*

## 1. INTRODUCTION

Image Contrast enhancement is representing as an essential pre-processing step for image segmentation [1] and [7]. The image contrast enhancement is said to be the differences between the visual properties of an objects that distinguish from another objects from the background of the image [8]. Contrast property is very important for images resolution where human eye can see the details of an image through the difference in contrast levels between these details. There are many reasons that cause low contrast images one of them is deprived brightness during image acquisition process [23].Many contrast enhancement techniques been found in the literature. Binarization is a simple technique that is used to enhance images in general [1] and to perform for the betterment to enhance the brightness of the image in particular. However, the main important aspects of Binarization techniques is to create histogram for analysis to get result that is caused by the uniform distribution of the image intensity [13]. On the other hand, using of histogram is also a useful way to enhance the contrast of the digital images. Menotti et., al., [10] proposed a solution using technique called as multi-histogram, which decomposes the image into sub-images and applying on each of them. Sun et., al., [7] proposed a contrast enhancement algorithm based on histogram specification by using the deferential information of input histogram. In addition to these two techniques, Yu and Bajaj [29] proposed contrast enhancement technique which computes that new intensity value of the pixels depending on three local statistical parameters; local minimum, local maximum and local average. This method estimated the statistical parameters of the histogram using maximum likelihood Estimation (MLE). Furthermore, it can be produced by desired histogram by shifting one or more than one node in order to have a reasonable separation. Therefore, histogram parameters and maximum likelihood estimation (MLE) is used to estimate the statistical parameters of the original histogram. This enables the method to enhance the image contrast that has been lost in the Binarization process. The structure of the paper is as follows - Section 2 presents related work along with maximum likelihood estimation. Section 3 describes the problem statement Section 4 and 5 show the formulation and proposed method. Section 6 represents the experimental results by applying the proposed approach on multi-modal images. Finally, section 7 outlines the main conclusion. This document describes, and is written to conform to, author guidelines for the journals of AIRCC series. It is prepared in Microsoft Word as a .doc document. Although other means of preparation are acceptable, final, camera-ready versions must conform to this layout. Microsoft Word terminology is used where appropriate in this document. Although formatting instructions may often appear daunting, the simplest approach is to use this template and insert headings and text into it as appropriate.

## 2. Related work

Many methods, such as global thresholding [21], local thresholding [14, 16, 4, 25], the statistical approaches [2, 6] used in the enhancement of the image brightness. In few articles, the entropy-based method [15] and feature extraction methods have been used. The edge-based methods [9] and multi-level classifiers [24] have been used for the enhancement and binarization of document images. The binarization being used due to the presence of gray-level blue over the image (i.e. shadows, non-uniform illumination, defects in some areas of the document), local thresholding methods are required to adapt techniques that can improvise the output of the images and text. In this paper we uses a simple maximum likelihood estimation (MLE) approach based on an assumption to estimate mean and variance along with Binarization method for the distribution of data pixels in each class. The MLE, compared to other classification methods like neural networks [27] and support vector machines [11], remains a popular classification tool. A major consideration in applying the MLE rule is to estimate optimal threshold along with binarization method and sample variances that can be used to generate condense bounds for the parameters. Researcher concentrated on the binarization method and list of papers presented related on that area (DIBCO'09[3]) part of ICDAR'09 conference. Among the binarization methods, authors [1] proposed a method to obtain optimal threshold by finding means of the histogram by using pixels. Its threshold value is calculated based on a gray level mean. The method has the advantage of being able to detect background regions and prevent distortion pixels from appearing on the output. Although the original method is time-consuming, various computationally low cost implementations, such as an integral image method [12] and a grid-based method [25], are available. An example of local and adaptive thresholding, which is based on the detection of is presented in [22]. Authors [5] have been proposed a new adaptive approach based on the combination of several state-of-the-art binarization methods, which considers the edge information as well. In this paper we propose an approach to enhance the brightness of the image by the MLE estimation.

## 3. Analytic Model

In document of image processing, a single-mode distribution for text and a multi-mode for background are usually used [17], [21]. This is because of the complex nature of the background on de-graded document images. Normally, a Gaussian model is used for each mode in these distributions [21],[17], [18]. For the case of the background distribution, we skip the assumption of having a multi-mode distribution, because the proposed method adapts a separate model to each pixel on the document image. Therefore, the distributions are highly local, and so in most cases the background distribution of a pixel is a single-mode one. This allows us to assume two histogram-based models, one for background and other foreground. The problem of binarization using MLE can be considered to estimate the statistical parameters values, to make the model spatially adaptive, it is assumed that each pixel on the image has its own binary (value). There are several decision rules that can be used. For example maximum likelihood estimation (MLE), minimum probability of error, maximum a posteriori, and Bayes risk decision [28]. Although MLE does not consider the a priori information on the classes), this cannot be considered a drawback. To find the suitable models for the particular class is a very difficult task with respect to the selection algorithm) because of the complexity Algorithm is high and variability of text and background on the historical documents. At the same time, the local nature of the method enables it to adapt to possible variations on the input image. In particular, variations in text and background intensity over the document image domain can be very local, and therefore a simple model for is not able to handle them. Considering the well-known behaviour of image and background information in the features space. Two simple, basic models can be used to model each of them: histogram-based model [20] and Gaussian models. In histogram-based models, the model of each class is estimated according to its probability density function [26]. Let us we

consider a set of observation forming a vector space $x = [x_1, ....., x_n]^\tau$ and $x_i \in X_i$

where $X_i$ is the random variable and $i \in I^{>0}$ set of natural number. Each random variable $X_i$ has a probability density function f (.) that has Unknown distribution with fixed parameter μ. The maximum likelihood estimation is given by

$$\mu_{ml}(x) = \arg Max\{f(x \mid \mu)\} ..................... (1)$$

Where,

$$f(x \mid \mu) = f(x_1 \mid \mu), f(x_2 \mid \mu), ..., f(x_n \mid \mu) .............................. (2)$$

is the likelihood function, The maximum likelihood estimation of $\mu$ is obtained as the solution to

$$(\partial/\partial u)f(x \mid \mu)_{\mu = \mu_{ml}} = 0 ..................................... (3)$$

and

$$(\partial^2/\partial^2 u)f(x \mid \mu)_{\mu = \mu_{\mu ml}} \prec 0 ......................... (4)$$

A bluer gray value document image, suffering from various de-gradation phenomena, such as bleed-through, dark background, or weak strokes, that can be expressed mathematically as:

$$\{u(x) \mid x = [i, j]^\tau \in \Omega \subseteq R^2\} .............................. (5)$$

The function mapped values range from 0 to 1, where a 0 value indicate the pixel is black, and a value of 1 means that the pixel is white. In addition, we assume that as a priori information, a rough binarization map of the image, u, R, is available. Although this approximate map may suffer from a large error, it is assumed that its recall value against the ground-truth binarization map is high. A high recall value is chosen to reduce the presence of interfering patterns (such as de-graded background, bleed-through, and show-through) on the map. The goal is to binarize the image, and separate them from the background and possible interfering patterns. In other words, a binarized map of u, R, will be the final output.

## 4. ALGORITHM DEVELOPMENT

The analytic model used to enhance the resolution of the bi-modal images by the using Binarization techniques. In this method we first, identify the statistical parameters of the original histogram of the image to estimate its statistical parameters value using maximum likelihood estimation (MLE). Then, it produced the desired histogram by shifting the first, the second, or the both nodes in order to have a reasonable separation to attain threshold.

## 4.1. The procedure to formulate the model

Algorithm is divided into different modules, Module one for histogram generation, module two for the parameters estimation and module  three for the modification of the histogram.

## 4.2. Algorithm for Module 1

Start

// Compute the Initial  histogram.

Step 1:  Read the Image.

Step 2: Chose an  origin $x_0$

// $B_j(x_0, h)$ denote the bin of length $h$ which is the element of a //

bin grid  starting at $x_0$

Step 3: Compute $B_j(x_0, h)$

Step 4: $B_j(x_0, h) = [x_0 + (j-1)h, x_0 + jh)$ Such that $j \in \mathrm{N}^{>0}$

// $\{x_i, i \in I^{>0}\}$ is the set of independent identical

// distributed sample with probability density function $f$

Step 5: select $x_i$ from $B_j(x_0, h)$

Step 6:  select $x$ from $B_j(x_0, h)$

Step 7: $I(x_i) = \{$ number of observation falling  into $B_j(x_0, h)$ $\}$

Step 8: $I(x) = \{$ localizing counts around $x$ $\}$

// compete $f^*$ that presents the shape of the histogram

Step 9: $f_h^*(x) = (nh)^{-1} \sum_{j \in Z} \sum_{i=1}^{n} I(x_i) I(x)$

End

## 4.3. Algorithm for Module 2

Start

// Apply the Binarization and Maximum Likelihood Estimation

//Techniques on the original histogram to get precise of Statistical

// information function $f^*$ for each mode.

Step 1: Noise Estimate

$$(\partial / \partial u) f(x \mid \mu)_{\mu = \mu_{ml}} = 0 \quad \text{Subject to condition}$$

$$(\partial^2 / \partial^2 u) f(x \mid \mu)_{\mu = \mu_{\mu ml}} \prec 0$$

// Estimate $\mu_{ml}(x)$

Step 2: $\theta = \mu_{ml}$

// Compute the statistical parameters

Step 3: Compute Standard Deviation,    mean Compute linear  process
Where, y = Observed data,   $\theta$ = Set of Modal
 //   parameters,      n = additive noise

Step 4:  $y = f^{*}\{\theta\} + n$

//    Separate the modes by shifting right, left thru Binarization
// techniques Max and Min is the largest and lowest gray
//     level values in the image

Step 5:   threshold == Value (max + min) / 2

//   Construct the desired histogram using data attain  from
//   Statistical parameters from Maximum Likelihood Estimation.

Step 6: Shift the means of mode 1 to mode $i$ to the left by $\mu_1$ - Min
Stept 7:  Shift the mean of the mode $i+1$ to the right by: (Max-$\mu_{i+1}$/(n-i)

Step 8: Shift the means of mode 1 to mode $i$ to the left by ($\mu_1$ − Min) / 2
and Shift the   mean of the mode $i+1$ to the right by (Max-$\mu_{i+1}$/2(n-i).
End
**Output:** The Enhanced Image

In this case, a very small range of pixels' intensities and shifting values will be very high as compared to bigger change or shift in the threshold value. This problem is due to the dependence on the maximum and minimum value of pixel's intensities in the image.

## 5. EXPERIMENTAL RESULTS

This section presents the experimental results of the proposed method. It displays these results by showing the image before and after the processing as well as the histograms for both images. The proposed method tested on multiple tri-modal gray-level images.

Fig. 1 (a) and Fig. 1 (b) show original and enhanced medical tri-modal images respectively. Fig. 1 (c) shows the original histogram and the desired histogram. This image enhanced by shifting the third mean to the right.

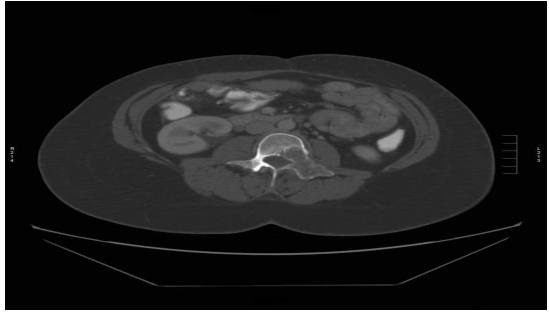

Figure 1 (a) Original Image

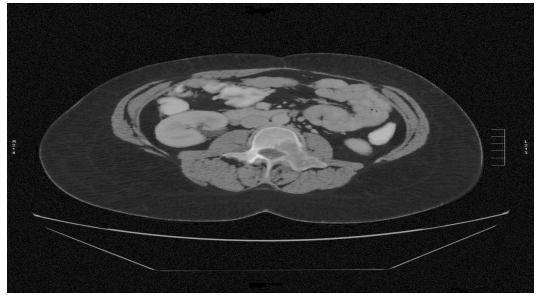

Figure 1 (b) Enhanced Image

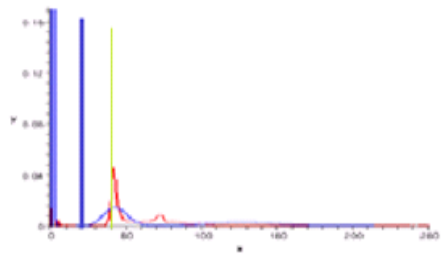

Figure 1 (c) Original (red) , desired (blue) histogram and threshold (green)

Fig. 2 (a) and Fig. 2 (b) show original and enhanced medical tri-modal images respectively. Fig. 2 (c) shows the original histogram and the desired histogram. This image is enhanced by shifting the third mean to the right.

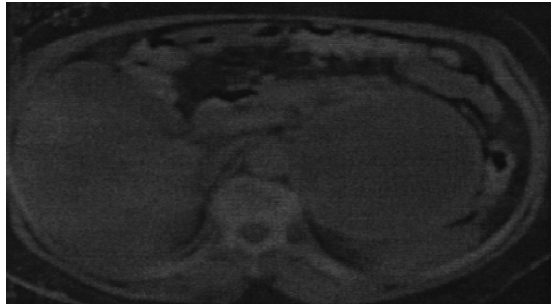

Figure 2 (a) Original Image.

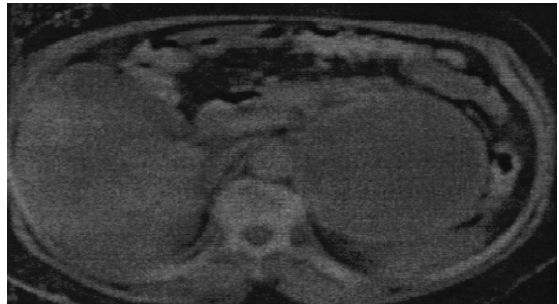

Figure 2 (b) Enhanced Image.

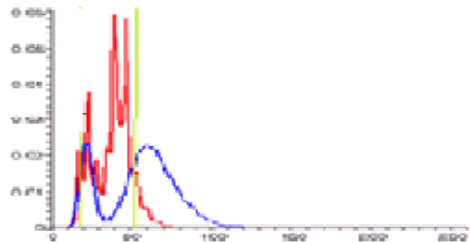

Figure 2(c) numerical values of the image

Figure 2 (a) Original Image. 2(b) Enhanced Image. 2(c) Original (red), desired (blue) histogram and threshold (green).

# 6. CONCLUSION

This paper presented a method to enhance the image using maximum likelihood estimation with Binarization techniques. The proposed method was based on the work of Aroop Mukherjee and Soumen Kanrar [1]. It aimed to improve and handle multi-modal images in addition to bi-modal ones. The proposed method of this paper was tested on various tri-modal gray-level images. The experimental results showed an improvement in the enhancement of these images using maximum likelihood estimation and Binarization techniques. Since the proposed method in this paper was tested only on bi-modal and tri-modal image, a future work of this method is recommended to complete testing on the remaining multi-modal images.

## Authors

Aroop Mukherjee is M.Phil (Computer Science), MCA, and M. Sc (Physics) from Vinayaka Mission University, India; National University (IGNOU), India and Jabalpur University India respectively.

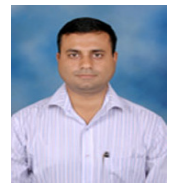

Currently, he is an editorial member cum reviewer for 5 international journals in the fields of Computer Science and IT Management (ISI Indexed) and also serving as reviewer for refereed conferences at international levels.

He is the member of IEEE and CSI.

Soumen Kanrar received the M.Tech. degree in computer science from Indian Institute of Technology Kharagpur India in 2000. Advanced Computer Programming RCC Calcutta India 1998. and MS degree in Applied Mathematics from Jadavpur University India in 1996. BS degree from Calcutta University India. Currently he is working as researcher at Vehere Interactive Calcutta India. Previously he had worked at King Saud University, Riyadh. He is the member of IEEE and CSI.

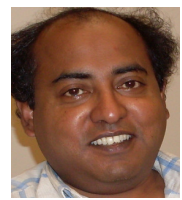

Soumen.kanrar@veheretech.com